\documentclass[10pt,twocolumn,letterpaper]{article}

\usepackage{cvpr}              

\usepackage{microtype}
\usepackage{graphicx}

\usepackage{booktabs} 
\usepackage{comment}

\usepackage{tikz}
\usepackage{adjustbox}

\usepackage{verbatim}

\usepackage[utf8]{inputenc} 
\usepackage[T1]{fontenc}    
\usepackage{url}            
\usepackage{nicefrac}       
\usepackage{xcolor}         

\usepackage{xspace}
\usepackage{amsmath,amssymb,amsfonts, mathtools}

\usepackage{censor}
\usepackage{textcomp}
\usepackage{nicefrac}       
\usepackage{caption}

\usepackage{algorithmic}
\usepackage{blindtext}
\usepackage[notransparent]{svg}

\usepackage{color,soul}
\usepackage{graphics}
\usepackage{listings}
\usepackage[toc,page]{appendix}
\usepackage{mwe}%
\usepackage{algorithm}

\usepackage{pgfplots}
\usepackage{siunitx}
\usepgfplotslibrary{groupplots}
\definecolor{darkred}{rgb}{0.55,0.0,0.0}
\pgfplotsset{compat=1.18}  
\usepackage{breakurl} 
\usepackage{xurl}

\usepackage{array,tabularx,multirow,url}
\newcolumntype{Y}{>{\raggedright\arraybackslash}X}

\usetikzlibrary{arrows.meta,positioning,fit,backgrounds,calc,shapes.geometric}

\definecolor{exactblue}{HTML}{235789}
\definecolor{forecastgreen}{HTML}{177E65}
\definecolor{protogold}{HTML}{9B6500}
\definecolor{riskred}{HTML}{AA3434}
\tikzset{box/.style={draw=exactblue,rounded corners=2pt,align=center,
  fill=exactblue!5,inner sep=5pt,font=\small},
  optbox/.style={box,draw=protogold,dashed,fill=protogold!6},
  flow/.style={-{Latex[length=2mm]},thick,draw=exactblue},
  optional/.style={flow,dashed,draw=protogold}}

\definecolor{mkvBlue}{HTML}{DCEBFF}
\definecolor{mkvBlueLine}{HTML}{3B6EA8}
\definecolor{mkvGreen}{HTML}{E3F4E6}
\definecolor{mkvGreenLine}{HTML}{4C8C5A}
\definecolor{mkvOrange}{HTML}{FFF0D6}
\definecolor{mkvOrangeLine}{HTML}{C47A1C}
\definecolor{mkvRed}{HTML}{FDE2E2}
\definecolor{mkvRedLine}{HTML}{B24A4A}
\definecolor{mkvGray}{HTML}{F4F4F4}
\definecolor{mkvDark}{HTML}{222222}

\definecolor{cvprblue}{rgb}{0.21,0.49,0.74}
\usepackage[pagebackref,breaklinks,colorlinks,allcolors=cvprblue]{hyperref}
\usepackage[all]{hypcap}
\pgfplotsset{compat=1.18}

\def\confName{CVPR}
\def\confYear{2027}

\title{EviCastKV: Query-Conditioned Multimodal KV Cache Compression}

\author{Arup Kumar Sarker, Mills Staylor, Gregor von Laszewski, Yue Cheng, Geoffrey Fox\\
University of Virginia \& Biocomplexity Institute and Initiatives, 
Charlottesville, VA, USA\\
{\tt\small \{djy8hg, qad5gv, thf2bn, mrz7dp, vxj6mb\}@virginia.edu} 
\and
Aymen Alsaadi, Shantenu Jha\\
Rutgers University \& Princeton Plasma Physics Laboratory, 
NJ, USA \\
{\tt\small \{aymen.alsaadi, shantenu.jha\}@rutgers.edu}
}   

\begin{document}
\maketitle
\begin{abstract}
Multimodal large language models produce large key--value (KV) caches, making compression a joint problem of evidence retention, physical allocation, and execution cost.  We present EviCastKV, a training-free approach to query-conditioned cache construction with exact-retention and shared-source configurations. Exact retention combines prompt-conditioned demand, value-sensitive inter-modal coverage, and routed planning. A shared capacity transaction packs selected original K/V entries under an initial physical-byte budget while preserving positions.
Across two LLaVA checkpoints, matched-capacity experiments show that EviCastKV improves COCO CIDEr from 107.84 to 111.48 over a local VL-Cache implementation at 40~MiB, with 19.65\% lower
mean request time. On MileBench at 16~MiB, it improves ROUGE-L by 2.21 points and Accuracy by 18.75 percentage points over a local LOOK-M implementation, with 38.6\% lower request time.
Prompt Top-K remains competitive in quality. EviCastKV's shared-source configuration quantizes a common projection input, protects selected entries exactly, and reconstructs layer-local operands. On 256 inspected COCO images, it uses 41.14/73.22~MiB initial/final decode-owned storage and scores 119.77 CIDEr. 
Together, these configurations characterize complementary quality--latency and storage trade-offs in query-conditioned multimodal cache construction.
\end{abstract}

\section{Introduction}
\label{sec:introduction}

Multimodal large language models (MLLMs) expand visual inputs into token prefixes processed with textual instructions \cite{liu2023llava,liu2024improved,bai2025qwen}. Their KV caches
retain representations for autoregressive decoding. Useful evidence is unevenly distributed: a small word, object relationship, or earlier frame may determine an answer despite extensive visual
redundancy. Compression must therefore reduce stored state without indiscriminately discarding evidence or introducing excessive planning and reconstruction costs.

SnapKV anticipates cache use from prompt observations \cite{li2024snapkv}; LOOK-M, PrefixKV, and VL-Cache address merging, importance-ranked allocation, and modality-aware selection \cite{wan2024lookm,wang2025prefixkv,tu2025vlcache}. FlashAttention and PagedAttention optimize attention execution and physical allocation \cite{dao2022flash,dao2023flash2,kwon2023paged}.
We ask: \emph{can evidence-aware selection improve quality at fixed physical capacity while keeping complete execution cost competitive?} Nominal retention and isolated decode
timings cannot establish this trade-off.

Three challenges govern this problem. \emph{Evidence selection} must estimate unknown decoding demand without allocating the budget only to repeated content. \emph{Physical feasibility}
must account for mandatory entries, padding, backing allocations, and original positional representations~\cite{su2021rope}. \emph{Execution efficiency} must control planning, construction, and reconstruction while accounting for generated-state growth.

We present \textbf{EviCastKV}, a training-free approach that couples query-conditioned evidence prioritization with position-correct cache construction (Figure~\ref{fig:evicastkv_design}). 
\textbf{In exact retention}, attention from a trailing window of $w$ prompt positions estimates demand; sparse inter-modal relationships, value-sensitive utility, and coverage guide routed selection. A physical capacity transaction reconciles the selected entries before packing their original K/V values. \textbf{In shared-source encoding}, prompt-attention observations determine exact protection, while a quantized common projection input represents the remaining prompt state. The runtime reconstructs layer-local K/V and losslessly codes its own generated history. The configurations examine two related decisions: which entries receive exact storage, and how the cache representation is materialized for decoding. Their executed policies and measurement protocols are identified in Section~\ref{sec:evaluation}.

Batch-one HF eager experiments cover COCO captioning and selected MileBench tasks \cite{lin2014coco,vedantam2015cider,song2024milebench}. At matched 40-MiB initial K/V backing, EviCastKV improves COCO CIDEr from 107.84 to 111.48 over local VL-Cache, with 19.65\% lower mean request time. At 16~MiB on MileBench, gains over local LOOK-M are 2.21 ROUGE-L points and 18.75 Accuracy percentage points, with 38.6\% lower time. Top-K remains competitive in quality. EviCastKV's shared-source configuration provides additional storage/runtime evidence where initial storage and mean audited decode peak decrease by 18.89\% and 10.03\%, respectively, compared to the nearest baseline Top-K adaptation.

\textbf{We contribute:}
(1) query-conditioned exact retention combining value-sensitive inter-modal coverage with routed planning; 
(2) position-correct cache construction through physically budgeted exact packing and shared-source encoding with protected entries and layer-local reconstruction; and
(3) configuration-specific evaluations of task quality, physical storage, and complete request cost, including matched-capacity gains and audited storage operating points.
\section{Background and Problem Formulation}
\label{sec:background}
\label{sec:problem}
\label{sec:background_problem}

\subsection{Multimodal Attention and Cache Growth}

Vision--language models combine visual representations and textual
instructions in an autoregressive decoder
\cite{liu2023llava,liu2024improved}. Prefill constructs layer-wise
keys and values; decoding reuses this state and appends newly
processed positions. For one attention head, query $q_t$ produces
\begin{equation}
 O_t=\operatorname{softmax}\!\left(
 \frac{q_tK_{\leq t}^{\top}}{\sqrt{d_h}}+M_t
 \right)V_{\leq t},
 \label{eq:evicast_attention}
\end{equation}
where $d_h$ is the head dimension and $M_t$ enforces causal and
validity constraints~\cite{vaswani2017attention}. Grouped-query
attention shares KV heads across query heads, so selection must
respect this mapping~\cite{ainslie2023gqa}. For batch size $b$,
$L$ layers, $H_{\mathrm{kv}}$ KV heads, $N$ cached positions,
and $s$ bytes per element, homogeneous dense K/V storage is
\begin{equation}
 B_{\mathrm{dense}}=2bLH_{\mathrm{kv}}Nd_hs.
 \label{eq:dense}
\end{equation}
Visual token expansion increases this persistent state, while
generation adds further positions. Consequently, equal output
lengths need not produce equal cache growth across checkpoints.

\subsection{Evidence and Representation Constraints}

Prompt observations provide a demand proxy~\cite{li2024snapkv},
but future decoder queries are unavailable during compression.
Attention weights alone omit value contribution. A chart mark
and its legend, for example, can provide complementary context;
independent pruning may remove one. Coverage motivates diversity
without guaranteeing preservation of every dependency.

Logical retention also differs from physical storage: padding,
preallocation, and views can retain more backing memory than
selected-token counts suggest. Compaction must preserve original
rotary positions~\cite{su2021rope}; moving position $p$ into
slot $j$ does not make it position $j$. Preserving retained
states does not preserve evicted information.

\subsection{Physically Constrained Cache Selection}

Let $\mathcal B$ contain blocks from the immutable prefill cache,
each associated with a layer, KV head, original positions, and
available modality metadata. Given mandatory blocks $\mathcal M$
and initial byte cap $C$, the primary exact-selection path constructs
a retained set $S$ satisfying
\begin{equation}
 \mathcal M\subseteq S\subseteq\mathcal B,\qquad
 B_{\mathrm{init}}\!\left(\operatorname{pack}(S)\right)\leq C.
 \label{eq:budget}
\end{equation}
Packing preserves selected original K/V entries. $B_{\mathrm{init}}$ counts unique active K/V backing, including padding and preallocation; the cap excludes construction workspace and subsequent append-only growth. The objective is to retain decoding-relevant evidence while controlling complete request cost. Section~\ref{sec:design} introduces utility and coverage surrogates, while task quality is measured empirically. Equation~\eqref{eq:budget} defines the exact-retention configuration. EviCastKV's shared-source configuration instead studies exact protection within an encoded representation; Section~\ref{sec:evaluation} specifies its realized storage and measurement scope.
\section{EviCastKV Design}
\label{sec:design}

\subsection{Overview and Design Contract}

\begin{figure*}[t]
\centering
\resizebox{\textwidth}{!}{%
\begin{tikzpicture}[
  font=\scriptsize,
  >=Latex,
  box/.style={
    draw=black!45,
    rounded corners=2pt,
    font=\scriptsize,
    text width=3.42cm,
    minimum height=1.12cm,
    align=center,
    inner sep=4pt
  },
  retention/.style={
    box,
    draw=blue!60!black,
    fill=blue!5
  },
  source/.style={
    box,
    draw=teal!60!black,
    fill=teal!5
  },
  flow/.style={
    ->,
    semithick,
    draw=black!65
  },
  stage/.style={
    font=\small\bfseries,
    align=center
  }
]

\draw[
  draw=black!30,
  rounded corners=3pt
] (-0.10,0.78) rectangle (16.10,-4.20);

\node[stage] at (1.90,0.43)
  {1. Prompt observations};
\node[stage] at (5.95,0.43)
  {2. Evidence policy};
\node[stage] at (10.00,0.43)
  {3. Cache construction};
\node[stage] at (14.05,0.43)
  {4. Decoding};


\node[
  anchor=west,
  font=\scriptsize\bfseries
] at (0.10,-0.18)
  {EviCastKV: exact-retention configuration
   (Tables~\ref{tab:coco-measured}--\ref{tab:mile40})};

\node[retention] (r1) at (1.90,-1.03)
  {Final $w$ prompt positions\\
   K/V descriptors\\
   Original positions};

\node[retention] (r2) at (5.95,-1.03)
  {Stable: prompt ranking\\
   Routed: inter-modal graph\\
   Value and coverage};

\node[retention] (r3) at (10.00,-1.03)
  {Capacity reconciliation\\
   Packed original K/V\\
   Initial physical-byte cap};

\node[retention] (r4) at (14.05,-1.03)
  {Attend to retained K/V\\
   Append generated state\\
   Preserve logical positions};

\draw[flow] (r1) -- (r2);
\draw[flow] (r2) -- (r3);
\draw[flow] (r3) -- (r4);


\draw[draw=black!18]
  (0.10,-1.88) -- (15.90,-1.88);

\node[
  anchor=west,
  font=\scriptsize\bfseries
] at (0.10,-2.13)
  {EviCastKV: shared-source configuration
   (Table~\ref{tab:storage-operating})};

\node[source] (s1) at (1.90,-2.98)
  {Normalized source $X_\ell$\\
   Prompt-attention observations\\
   Original positions};

\node[source] (s2) at (5.95,-2.98)
  {Windowed attention ($w$)\\
   + full-prompt prior (1:1)\\
   Choose exact protection};

\node[source] (s3) at (10.00,-2.98)
  {Three-bit source\\
   Exact overlays + corrections\\
   Lossless generated history};

\node[source] (s4) at (14.05,-2.98)
  {Layer-local reconstruction\\
   Frozen $W^K,W^V$ + RoPE\\
   Attend; reuse buffers};

\draw[flow] (s1) -- (s2);
\draw[flow] (s2) -- (s3);
\draw[flow] (s3) -- (s4);

\node[align=center] at (8.00,-3.91)
  {Cross-cutting requirements: original positions;
   validated publication; physical-byte accounting.};

\end{tikzpicture}%
}
\caption{EviCastKV's evaluated cache-construction configurations.
Exact retention uses routed inter-modal evidence selection and
physical-capacity reconciliation. Shared-source encoding uses
attention-based exact protection and layer-local reconstruction.
Rows are alternative configurations, not consecutive stages;
measurement scopes and execution policies are defined per protocol. Here, $w$ denotes the prompt-observation window length.}
\label{fig:evicastkv_design}
\end{figure*}
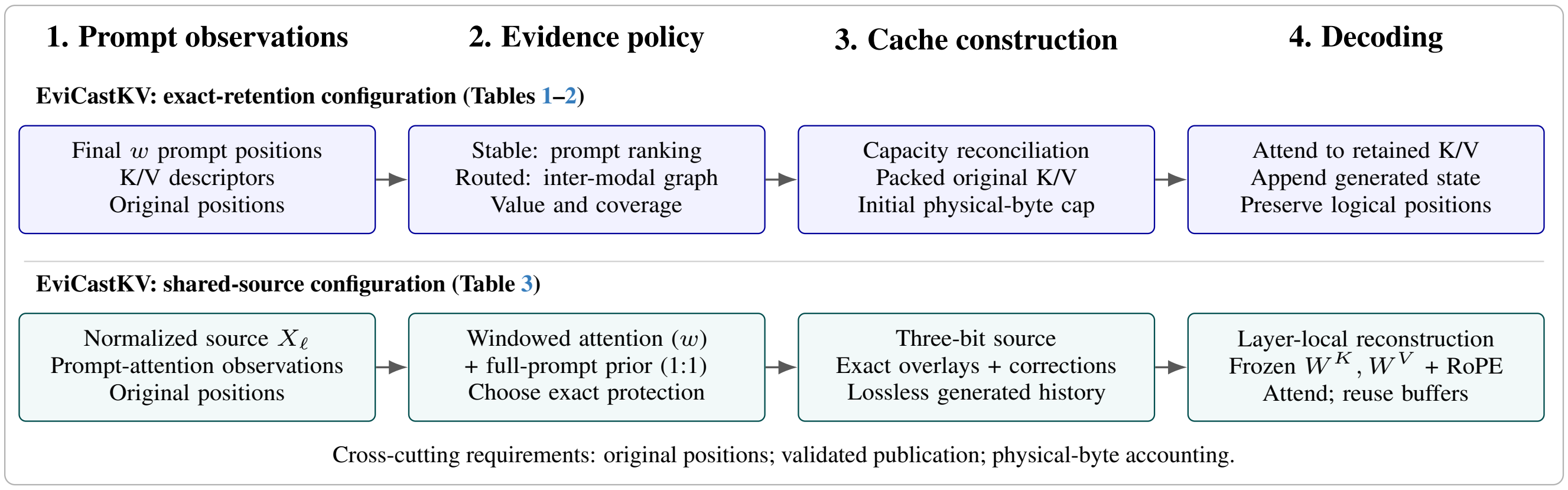

EviCastKV has two evaluated cache-construction configurations (Figure~\ref{fig:evicastkv_design}):
Exact retention  uses routed graph-based selection and exact packing, whereas shared-source encoding combines prompt-attention-based exact protection with a quantized projection source. The source configuration does not execute the graph, utility, or coverage selector of Sections~\ref{sec:design_evidence}--\ref{sec:design_routing}.


\subsection{Forecasting Demand from Prompt Queries}
\label{sec:design_forecast}

Following the observation-window principle of
SnapKV~\cite{li2024snapkv}, EviCastKV uses attention
queries from the final $w$ positions of the current prompt
as a proxy for subsequent demand, where $w$ denotes the
prompt-observation window length. Query heads are mapped to their owning KV heads, including grouped-query mappings~\cite{ainslie2023gqa};
aggregation remains within each layer/KV-head group. For block $b_i$,
\begin{equation}
 a_i=\sum_{t\in b_i}\sum_{j\in\mathcal W_{\ell h}}
       \omega_j A_{\ell h}(j,t),
 \qquad \sum_{j\in\mathcal W_{\ell h}}\omega_j=1,
 \label{eq:evicast_demand}
\end{equation}
where $A_{\ell h}(j,t)$ is position-compatible attention and $\mathcal W_{\ell h}$ contains query observations from those $w$ positions mapped to layer $\ell$ and KV head $h$. Attention uses
the original causal mask and compatible rotary transforms~\cite{su2021rope};
shortlisting must not renormalize attention over fewer keys. This
forecaster estimates cache demand, not future query vectors or factual
relations. The primary path uses no draft rollout, oracle future queries,
or benchmark references.

\subsection{Sparse Evidence and Value-Sensitive Utility}
\label{sec:design_evidence}

Within routed groups, eligible blocks form a sparse graph $G=(U,E)$.
Pooled K/V descriptors and prompt-attention co-activation provide
relationships; spatial, temporal, region, and layout metadata contribute
only when available. Edges respect layer/KV-head ownership, permitted
modality pairs, and positional spans. An edge denotes a potential
relationship, not semantic equivalence. Compatible sources are combined,
duplicates coalesced, and the final graph is limited to $d_g=8$ neighbors
per node. Storage is $O(|U|d_g)$, but chunked exhaustive search can still
require quadratic comparisons; routing and shortlisting restrict its
input size.

Let $v_i$ denote value contribution and $n_i$ neighbor-relative novelty.
Their raw proxies are $a_i\|\bar v_i\|_2$ and
$1-\max_{j\in\mathcal N(i)}\max\{0,\cos(\bar v_i,\bar v_j)\}$,
where $\bar v_i$ is a pooled value descriptor and $\mathcal N(i)$
contains its graph neighbors. Each signal is normalized to unit sum
within its layer/head group when its sum is nonzero. Reusing the symbols
for these normalized signals, the score is
\begin{equation}
 u_i=\lambda_q a_i-\lambda_v v_i-\lambda_o n_i.
 \label{eq:utility}
\end{equation}
The signed convention is retained: negative $\lambda_v$ and $\lambda_o$
reward contribution and novelty. These are ranking surrogates, not
measures of semantic necessity. For nonnegative edge weights $w_{ij}$ and modality labels $m(i)$,

\begin{align}
 C_g(S)&=\frac{1}{|U|}\sum_{j\in U}\max_{i\in S}w_{ij},
 \label{eq:evicast_graph_coverage}\\
 C_m(S)&=\frac{|\{m(i):i\in S\}|}{|\{m(i):i\in U\}|},
 \label{eq:evicast_modality_coverage}\\
 F(S)&=\sum_{i\in S}u_i-\lambda_g C_g(S)-\lambda_m C_m(S).
 \label{eq:objective}
\end{align}
Empty-set coverage is zero. Facility location gives smaller marginal
rewards to already-represented neighborhoods; it does not explicitly
penalize duplicate selection or require complementary pairs. Together,
the terms encourage value-weighted diversity and modality coverage.
For fixed nonnegative utilities and $\lambda_g,\lambda_m\leq0$, the
modular and coverage terms make $F$ monotone submodular. The classical
cardinality-constrained greedy guarantee~\cite{nemhauser1978} does not
certify the routed, padded, byte-constrained executor or task quality.

\textbf{Selection example:} Consider equal-cost candidates $a,b,c$: $a$ and $b$ represent
similar visual content, while $c$ represents distinct text.
After selecting $a$, coverage can favor $c$ when
$\Delta F(c\mid\{a\})>\Delta F(b\mid\{a\})$, where
$\Delta F(i\mid S)=F(S\cup\{i\})-F(S)$.
This illustrates diversity-sensitive selection, not guaranteed
preservation of an answer-critical pair.

\subsection{Routed Planning and Physical Construction}
\label{sec:design_routing}
\label{sec:design_capacity}
\label{sec:design_positions}

A deterministic policy partitions layer/KV-head groups into
stable and routed sets. Stable heads use vectorized prompt
ranking; routed heads apply the evidence objective to restricted
candidates. The primary configuration routes $1/256$ of groups
and shortlists 8\% of eligible nonmandatory blocks, expanded with
compatible neighbors. Mandatory entries are reserved first.
Vectorized coverage selection uses two correction rounds rather
than solving~\eqref{eq:objective} exactly. 
Both paths retain original position identities and share the physical
transaction in Algorithm~\ref{alg:exact}.

\begin{algorithm}[t]
\caption{EviCastKV exact-retention configuration}
\label{alg:exact}
\begin{algorithmic}[1]
\REQUIRE Prefill cache $\mathcal B$, prompt observations,
mandatory set $\mathcal M$, initial byte cap $C$, observation-window length $w$
\STATE Build block descriptors and original-position maps
\STATE Estimate demand from the final $w$ prompt positions
\STATE Partition layer/KV-head pairs into stable and routed sets
\STATE Select stable entries by vectorized prompt ranking
\STATE Build routed graphs and select by utility and coverage
\STATE Reconcile both paths, mandatory entries, and padding
\STATE Fill attainable capacity; reject infeasible mandatory state
\STATE Pack unchanged K/V entries and validate masks and bytes
\STATE Publish the cache; decode using original logical positions
\end{algorithmic}
\end{algorithm}

Both selections enter one capacity transaction under
\eqref{eq:budget}, including per-head minima, padding, and
preallocation. Retained lengths $(5,5)$ and $(9,1)$ select ten
entries but require ten and eighteen padded slots, respectively;
global top-$K$ therefore cannot replace physical reconciliation.
The transaction records attainable allocation and reports
mandatory infeasibility. Different allocation granularities
require a common feasible target or an explicit discrepancy.

Selected original entries are packed in position order with
validity masks. Their RoPE representations remain unchanged
\cite{su2021rope}, and decoding advances from the original
prefix length rather than the packed length. Publication follows
validation of tensors, positions, masks, and bytes. The initial
cap does not constrain subsequent append-only growth.

\subsection{Shared-Source Representation}
\label{sec:design_storage}

In EviCastKV's shared-source configuration, an encoded normalized attention input $X_\ell$ replaces the persistent storage of both projected arrays. For the supported full-MHA geometry, decoding the source gives
\begin{equation}
 \widetilde K_\ell=\operatorname{RoPE}
   (\widehat X_\ell W_\ell^K;P),\qquad
 \widetilde V_\ell=\widehat X_\ell W_\ell^V,
 \label{eq:evicast_source_reconstruction}
\end{equation}
using frozen weights and original positions. Grouped affine
three-bit storage uses groups of 128. Exact overlays protect regions chosen by an equally weighted combination of attention over the final $w$ prompt positions and a full-prompt prior, rather than
the graph objective. All prompt positions remain represented exactly or approximately. Numerical patches target native projection of the same decoded source, not unrestricted FullKV. Generated history seals 16-token chunks, retains an unfinished tail, and uses exact corrections to preserve its own BF16 K/V words. Layer-local reconstruction trades computation for persistent storage; codes, overlays, corrections, metadata, and live buffers are counted within their declared scopes.

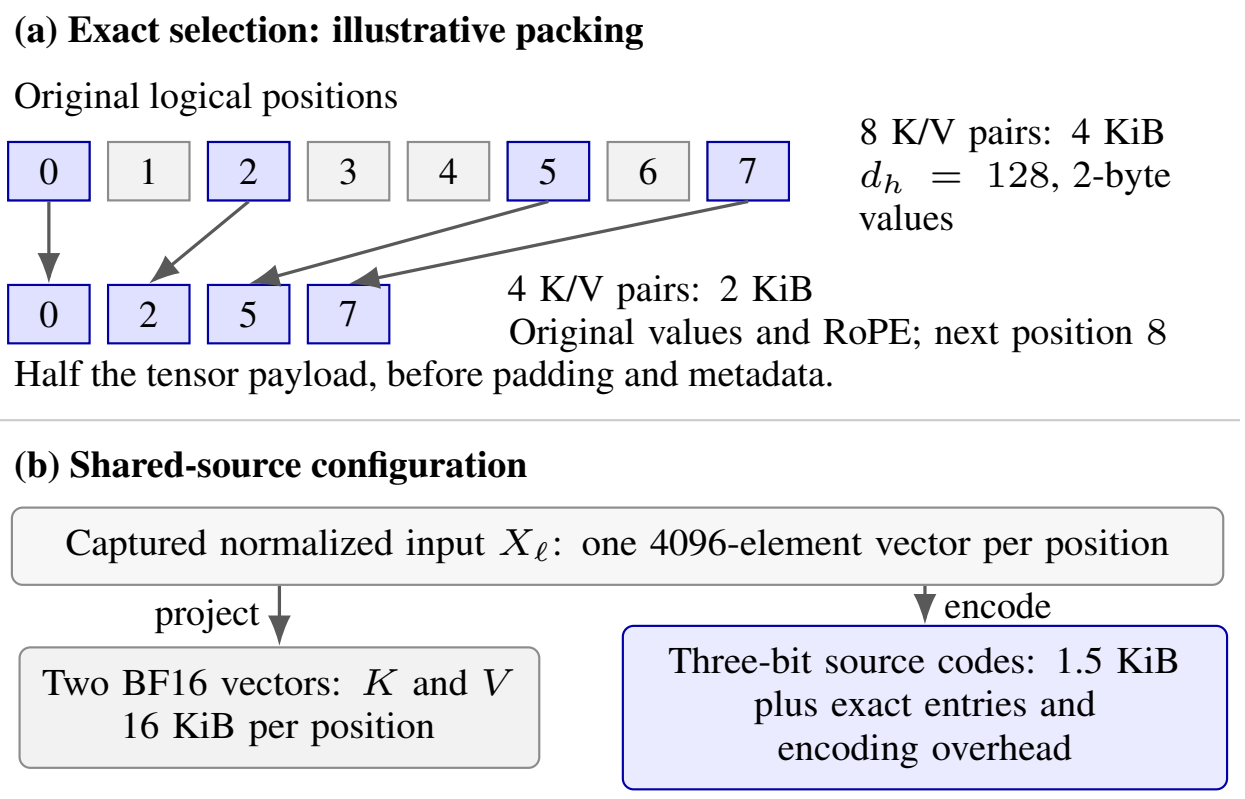
\begin{figure}[t]
\centering
\resizebox{\columnwidth}{!}{%
\begin{tikzpicture}[
  font=\scriptsize,
  >=Latex,
  slot/.style={
    draw=black!45,
    fill=black!5,
    minimum width=0.53cm,
    minimum height=0.38cm,
    inner sep=1pt
  },
  kept/.style={
    slot,
    draw=blue!65!black,
    fill=blue!12
  },
  box/.style={
    draw=black!45,
    rounded corners=2pt,
    align=center,
    inner sep=4pt
  },
  flow/.style={
    ->,
    semithick,
    draw=black!65
  }
]


\node[
  anchor=west,
  font=\scriptsize\bfseries
] at (0,0)
  {(a) Exact selection: illustrative packing};

\node[anchor=west] at (0,-0.43)
  {Original logical positions};

\foreach \p in {0,...,7}{
  \node[slot] (src\p)
    at ({0.35+0.65*\p},-0.90) {\p};
}

\foreach \p in {0,2,5,7}{
  \node[kept] at (src\p.center) {\p};
}

\foreach \p/\j in {0/0,2/1,5/2,7/3}{
  \node[kept] (dst\j)
    at ({0.35+0.65*\j},-1.83) {\p};
  \draw[flow] (src\p.south) -- (dst\j.north);
}

\node[
  anchor=west,
  align=left,
  text width=2.45cm
] at (5.50,-0.92)
  {8 K/V pairs: 4 KiB\\
   $d_h=128$, 2-byte values};

\node[
  anchor=west,
  align=left,
  text width=4.60cm
] at (3.22,-1.83)
  {4 K/V pairs: 2 KiB\\
   Original values and RoPE; next position $8$};

\node[anchor=west] at (0,-2.24)
  {Half the tensor payload, before padding and metadata.};

\draw[black!20] (0,-2.52) -- (8.10,-2.52);


\node[
  anchor=west,
  font=\scriptsize\bfseries
] at (0,-2.83)
  {(b) Shared-source configuration};

\node[
  box,
  fill=black!3,
  text width=7.60cm
] (x) at (4.05,-3.34)
  {Captured normalized input $X_\ell$:
   one 4096-element vector per position};

\node[
  box,
  fill=black!5,
  text width=3.10cm
] (dense) at (1.85,-4.39)
  {Two BF16 vectors: $K$ and $V$\\
   16 KiB per position};

\node[
  box,
  draw=blue!65!black,
  fill=blue!8,
  text width=3.65cm
] (codes) at (6.05,-4.39)
  {Three-bit source codes: 1.5 KiB\\
   plus exact entries and encoding overhead};

\draw[flow]
  (x.south -| dense.north)
  -- node[left]{project}
  (dense.north);

\draw[flow]
  (x.south -| codes.north)
  -- node[right]{encode}
  (codes.north);

\end{tikzpicture}%
}
\caption{Distinct storage mechanisms. (a) Packing halves one head's
K/V payload without renumbering retained positions. (b) In full MHA,
one encoded source replaces two persistent projections. These are
illustrative payload calculations, excluding metadata, overlays, and
scratch; neither panel asserts unchanged answer quality.}
\label{fig:evicastkv_savings_example}
\end{figure}

\textbf{Storage example: } Figure~\ref{fig:evicastkv_savings_example} illustrates the
payload opportunity. Packing four of eight K/V pairs with
$d_h=128$ and two-byte elements reduces payload from 4 to
2~KiB while preserving original positions. In full MHA,
three-bit codes for a 4096-element source occupy 1.5~KiB,
versus 16~KiB for two BF16 projections. These calculations
exclude encoding overhead and temporary buffers; quantization
also introduces approximation.

\section{Implementation}
\label{sec:implementation}
EviCastKV uses Python/PyTorch with model-specific adapters
and Hugging Face eager attention. The following paragraphs
describe the exact-retention and shared-source execution
configurations shown in Figure~\ref{fig:evicastkv_design}.

\textbf{Runtime and publication.}
Adapters expose prefill, query capture, cache extraction, and
one-token decoding while preserving rotary positions and
query-to-KV-head mappings~\cite{su2021rope,ainslie2023gqa}.
Contiguous prompt scores, segmented graph operations, descriptor
reuse, and layer-level transfers implement the routed policy;
stable heads bypass graph construction. A shared transaction
validates mandatory membership, positions, masks, and allocated
bytes before publishing the packed cache. Rejected proposals
remain failures. An explicit loop replaces
\texttt{model.generate()}: the first prediction uses original
prefill logits, so first-token agreement does not validate
compressed-cache behavior.

\paragraph{Shared-source execution:}
The shared-source configuration uses the pinned LLaVA-1.5-7B full-MHA configuration
with BF16 model execution. Forward hooks capture normalized
attention inputs and original rotary phases. Custom GPU kernels,
cuBLASLt projections, and reusable request-private descriptors
reconstruct layer-local operands. Validated packets replace
superseded original owners; reconstructed K/V is not retained
densely across all layers. Eligible attention segments use
request-private CUDA replay, with capture and retirement inside
the request. The capability is offered to all compressed adaptations,
but engages only for EviCastKV’s shared-source configuration under the measured layouts.
Consequently, latency reflects representation, layout, and execution
policy together, not an isolated codec benefit.

\paragraph{Where the execution cost can decrease: }
In the exact-selection path, routing limits graph construction,
not transformer execution: stable heads use tensorized prompt
ranking, while only routed groups incur sparse-graph and coverage
work. Reusing descriptors and reconciling selections before
materialization avoids constructing independent full-budget caches
for the two paths. Selection and packing are paid after prefill;
subsequent forwards consume the published representation. At matched initial/final K/V backing on MileBench(40~MiB), EviCastKV/Top-K mean times are 10.50/11.63~s for planning/construction, 2.77/2.79~s for decoding, and 16.71/18.03~s overall. The larger difference is in construction; these measurements do not isolate contributions from
routing, graph construction, or descriptor reuse.

\begin{figure}[t]
\centering
\resizebox{\columnwidth}{!}{%
\begin{tikzpicture}[
  font=\scriptsize,
  >=Latex,
  box/.style={
    draw=black!45,
    rounded corners=2pt,
    align=center,
    inner sep=4pt,
    minimum height=0.66cm
  },
  flow/.style={
    ->,
    semithick,
    draw=black!65
  }
]

\node[
  box,
  draw=blue!65!black,
  fill=blue!9,
  text width=7.30cm
] (persistent) at (4.05,0)
  {Persistent state across layers:
   encoded prompt + protected entries\\
   sealed history + unfinished tail + corrections + metadata};

\node[
  box,
  fill=orange!10,
  text width=1.95cm
] (lease) at (1.20,-1.30)
  {Reconstruct layer $\ell$\\
   Temporary $K_\ell,V_\ell$};

\node[
  box,
  fill=green!8,
  text width=1.95cm
] (attention) at (4.05,-1.30)
  {Attention\\
   Consume K/V\\
   Eligible replay};

\node[
  box,
  fill=black!4,
  text width=1.95cm
] (release) at (6.90,-1.30)
  {Return lease\\
   Reuse buffer};

\draw[flow]
  (persistent.south -| lease.north)
  -- (lease.north);

\draw[flow] (lease) -- (attention);
\draw[flow] (attention) -- (release);

\draw[flow]
  (release.south)
  -- (6.90,-2.12)
  -- (1.20,-2.12)
  -- (lease.south);

\node[
  align=center,
  text width=7.60cm
] at (4.05,-2.50)
  {Next layer: reconstruct new operands in reused storage.};

\end{tikzpicture}%
}
\caption{EviCastKV's shared-source reconstruction lifetime. The decode audit counts persistent packets and live temporary overlap; it does not certify whole-request memory. Replay setup and retirement are timed. This is a separate execution path, not a replacement for routed
evidence selection.}
\label{fig:evicastkv_runtime_lifetime}
\end{figure}
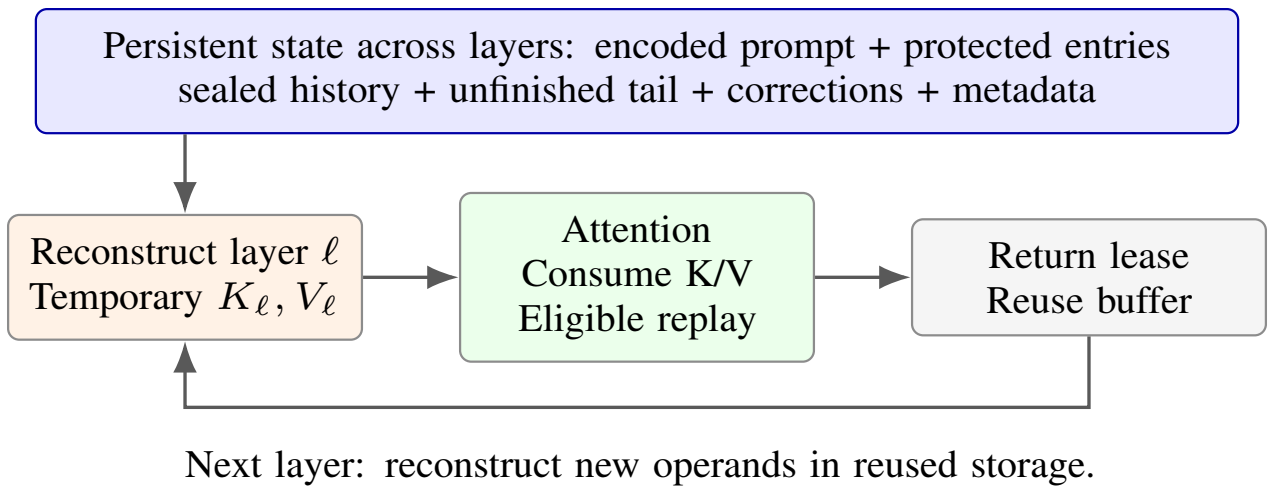

\paragraph{Reconstruct, consume, and reuse: }
Figure~\ref{fig:evicastkv_runtime_lifetime} shows EviCastKV's shared-source execution. Persistent encoded state supplies layer-local operands through a bounded lease. Buffers are reused after attention consumes their contents; each layer reconstructs its own K/V. Validated publication retires superseded original owners, avoiding a hidden full-cache copy. Shared-source encoding uses 41.14~MiB initially versus FullKV's 301.50~MiB; its mean audited decode peak is
83.52~MiB. Median times are 6.55~s versus Top-K's 7.50~s and FullKV's 5.85~s. Only shared-source layouts engage replay, confounding attribution; these BF16 measurements do not represent optimization of the FP16 selector.

\paragraph{Correctness boundaries:}
For exact selection, retention-one checks compare at least
16 greedy tokens against ordinary generation and cache
extraction/reinjection, requiring identical tokens and maximum
absolute logit error $10^{-4}$. These checks validate identity
transport, not arbitrary pruned layouts. The shared-source configuration instead checks reconstruction against native projection of the same decoded source, using absolute tolerance $2^{-6}$ and relative tolerance $0.01$ under the pinned dtype predicate. Protected
entries and generated-history corrections have bitwise contracts;
history exactness refers to the variant's own forward passes.
Neither numerical contract establishes answer-quality preservation
relative to unrestricted FullKV. 

\paragraph{Physical accounting, instrumentation and provenance:}
For unique backing allocations $\mathcal A(t)$ within a declared
measurement scope and observation times $\mathcal T$, we record
\begin{align}
 B_{\mathrm{active}}(t)
   &=\sum_{a\in\mathcal A(t)}\operatorname{nbytes}(a),
 \label{eq:impl_live_storage}\\
 B_{\mathrm{peak}}
   &=\max_{t\in\mathcal T}B_{\mathrm{active}}(t).
 \label{eq:impl_boundary_peak}
\end{align}
Backing allocations are deduplicated by device and storage
identity; padding and preallocation are charged in full.
Exact-selection accounting measures active K/V, with metadata
and planner workspace separate. The shared-source audit also
counts encoded metadata and observed reconstruction/append
overlap, but excludes prefill/construction transients, ordinary
attention activations, and vendor-internal workspace.
Neither scope measures whole-request GPU memory.

CUDA events delimit device phases; synchronized wall time
captures complete requests. Memory-audit passes remain separate
from headline timing. Immutable manifests bind source,
configuration, inputs, model/data revisions, scorers, software,
and hardware. Failures remain recorded, and retries do not
increase sample counts. Evaluation specifies each protocol's
timing boundaries and aggregation.

\section{Evaluation}
\label{sec:evaluation}

We evaluate EviCastKV through two cache-construction
configurations. The exact-retention configuration uses
routed evidence selection and physical-capacity reconciliation
(Tables~\ref{tab:coco-measured}--\ref{tab:mile40} and
Figure~\ref{fig:curves}). The shared-source configuration
uses attention-based exact protection and encoded source
reconstruction (Table~\ref{tab:storage-operating}).
Each comparison retains its checkpoint, precision,
generation schedule, and measurement scope.

\subsection{Experimental Configurations and Controls}
\label{sec:eval_protocol}

\paragraph{Primary study:}
Experiments use A100-SXM4-80GB hardware, 12 CPU cores per GPU,
batch-one HF eager execution, PyTorch~2.8.0, and
Transformers~4.55.2. We evaluate LLaVA-v1.6-Mistral-7B on 256 COCO
images and LLaVA-1.5-7B on 192 MileBench examples
\cite{liu2023llava,liu2024improved,lin2014coco,song2024milebench}.
Both use FP16 and the 64-token greedy-generation protocol. Initial
budgets are $\{10,20,40,80\}$~MiB for COCO and
$\{16,20,28,40,80\}$~MiB for MileBench and prompt-observation window of $w=16$ in the reported experiments. The latter covers ActionPrediction, ActionSequence, CLEVR-Change, and Spot-the-Diff: 96 temporal examples contribute Accuracy and 96 semantic examples
contribute ROUGE-L. 


\paragraph{Comparators and statistics:}
Prompt Top-K appears in both primary protocols, alongside local
VL-Cache on COCO and LOOK-M on MileBench
\cite{tu2025vlcache,wan2024lookm}. R denotes reimplementation;
VL-Cache omits its specialized fused kernels. Within each protocol,
compressed methods share processed inputs, precision, protected
positions, one-token granularity, generation, and realized K/V storage.
Five warmups precede one measured request trial; reported latencies
are means. Corpus CIDEr and category-aggregated ROUGE-L remain separate
\cite{vedantam2015cider,lin2004rouge}. Primary paired bootstrap
intervals use 10,000 resamples and 95\% confidence
\cite{efron1979bootstrap}; a zero-crossing interval does not establish
equivalence. COCO and the additional MileBench cohort have prior
protocol use; the primary MileBench cohort was exclusion-reviewed.

\paragraph{Memory scopes:} Tables~\ref{tab:coco-measured} and~\ref{tab:mile40} count
initial/final active K/V backing, including padding and
preallocation; metadata and workspace are separate.
Table~\ref{tab:storage-operating} counts decode-owned storage,
including encoded metadata and observed reconstruction/append
overlap. Its mean peak averages per-request audited maxima.
The original boundary maxima are not equivalent to this audit,
and neither scope measures whole-request GPU memory.

\subsection{COCO: Quality at Matched Physical Capacity}
\label{sec:eval_coco}

\begin{table}[t]
\centering
\caption{COCO exact selection: LLaVA-v1.6-Mistral-7B, FP16,
256 images, 64-token protocol. Memory columns are mean active K/V
backing, not transient-inclusive peaks. Times are means; FullKV
latency was not measured. Bold marks the best reported quality or E2E
point estimate.}
\label{tab:coco-measured}
\begingroup
\footnotesize
\setlength{\tabcolsep}{2pt}
\begin{tabular*}{\columnwidth}{@{\extracolsep{\fill}}lrrrrr@{}}
\toprule
& & \multicolumn{2}{c}{K/V backing (MiB)}
  & \multicolumn{2}{c}{Time (s)}\\
\cmidrule(lr){3-4}\cmidrule(l){5-6}
Method & CIDEr & Initial & Final & Decode & E2E\\
\midrule
FullKV
  & 111.24
  & 285.61
  & 293.481
  & ---
  & ---\\
Prompt Top-K
  & \textbf{111.59}
  & 40.00
  & 47.875
  & 2.95
  & 17.22\\
VL-Cache
  & 107.84
  & 40.00
  & 47.875
  & 3.12
  & 20.75\\
EviCastKV
  & 111.48
  & 40.00
  & 47.875
  & 2.95
  & \textbf{16.68}\\
\bottomrule
\end{tabular*}
\endgroup
\end{table}

At identical 40-MiB initial and 47.875-MiB final K/V storage,
EviCastKV improves CIDEr over VL-Cache from 107.84 to 111.48:
a paired gain of $3.64$, interval $[0.53,7.06]$
(Table~\ref{tab:coco-measured}). Mean request time decreases by
19.65\%, from 20.75 to 16.68~s; the paired absolute reduction has
interval $[3.92,4.24]$~s. Thus, the measured quality and request-time
gains do not rely on assigning EviCastKV a larger cache.

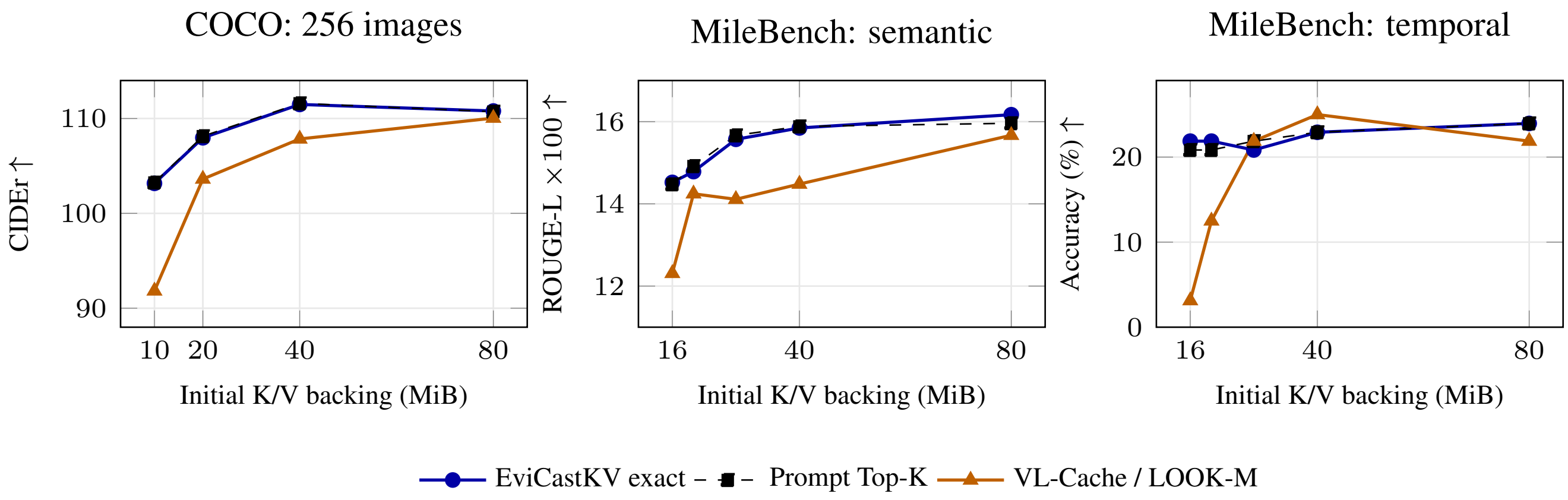
\begin{figure*}[t]
\centering
\begingroup
\pgfplotsset{every axis/.style={}}
\resizebox{\textwidth}{!}{%
\begin{tikzpicture}
\begin{groupplot}[
  group style={
    group name=evqualityscope,
    group size=3 by 1,
    horizontal sep=10mm
  },
  width=0.30\textwidth,
  height=3.8cm,
  xlabel={Initial K/V backing (MiB)},
  grid=major,
  grid style={draw=gray!18},
  tick label style={font=\scriptsize},
  label style={font=\scriptsize},
  title style={font=\small},
  legend style={font=\scriptsize,draw=none},
  legend columns=3,
  every axis plot/.append style={/tikz/mark size=1.6pt}
]

\nextgroupplot[
  title={COCO: 256 images},
  ylabel={CIDEr $\uparrow$},
  xtick={10,20,40,80},
  ymin=88,
  ymax=114,
  legend to name=evicastscopelegend
]
\addplot[blue!65!black,thick,mark=*] coordinates {
  (10,103.162617)
  (20,107.965110)
  (40,111.480002)
  (80,110.789294)
};
\addlegendentry{EviCastKV exact}

\addplot[black,dashed,mark=square*] coordinates {
  (10,103.240902)
  (20,108.127981)
  (40,111.592886)
  (80,110.715508)
};
\addlegendentry{Prompt Top-K}

\addplot[orange!75!black,thick,mark=triangle*] coordinates {
  (10,91.8250257)
  (20,103.616712)
  (40,107.844325)
  (80,110.038763)
};
\addlegendentry{VL-Cache / LOOK-M}

\nextgroupplot[
  title={MileBench: semantic},
  ylabel={ROUGE-L $\times100$ $\uparrow$},
  xtick={16,40,80},
  ymin=11,
  ymax=17
]
\addplot[blue!65!black,thick,mark=*] coordinates {
  (16,14.5227206)
  (20,14.7849767)
  (28,15.5700569)
  (40,15.8465021)
  (80,16.1689635)
};

\addplot[black,dashed,mark=square*] coordinates {
  (16,14.4748850)
  (20,14.9177124)
  (28,15.6657364)
  (40,15.8832763)
  (80,15.9607500)
};

\addplot[orange!75!black,thick,mark=triangle*] coordinates {
  (16,12.3091085)
  (20,14.2451096)
  (28,14.1109848)
  (40,14.4815875)
  (80,15.6691480)
};

\nextgroupplot[
  title={MileBench: temporal},
  ylabel={Accuracy (\%) $\uparrow$},
  xtick={16,40,80},
  ymin=0,
  ymax=29
]
\addplot[blue!65!black,thick,mark=*] coordinates {
  (16,21.875)
  (20,21.875)
  (28,20.8333333)
  (40,22.9166667)
  (80,23.9583333)
};

\addplot[black,dashed,mark=square*] coordinates {
  (16,20.8333333)
  (20,20.8333333)
  (28,21.875)
  (40,22.9166667)
  (80,23.9583333)
};

\addplot[orange!75!black,thick,mark=triangle*] coordinates {
  (16,3.125)
  (20,12.5)
  (28,21.875)
  (40,25)
  (80,21.875)
};

\end{groupplot}

\node[anchor=north] at
  ($(evqualityscope c1r1.south)!0.5!
    (evqualityscope c3r1.south)+(0,-10mm)$)
  {\pgfplotslegendfromname{evicastscopelegend}};

\end{tikzpicture}%
}
\endgroup
\caption{EviCastKV quality at matched initial K/V backing. Each MileBench metric covers 96 examples. Orange denotes VL-Cache on COCO and LOOK-M on MileBench. Lines connect all measured capacities.}
\label{fig:curves}
\end{figure*}

Figure~\ref{fig:curves} shows stronger quality gains over this port
at tight capacities: $+11.34$ CIDEr at 10~MiB, interval
$[6.55,16.49]$, and $+4.35$ at 20~MiB, interval $[0.01,8.90]$.
The $+0.75$ gain at 80~MiB is unresolved. Top-K remains competitive:
at 40~MiB, EviCastKV differs by $-0.11$ CIDEr, interval
$[-0.63,0.31]$, with 3.14\% lower mean request time. Against FullKV,
initial/final K/V storage is 86.0\%/83.7\% smaller, but the quality
difference $+0.24$, interval $[-3.74,4.51]$, does not establish
quality preservation.


\subsection{MileBench: Quality at Matched Capacity}
\label{sec:eval_milebench}

At 16~MiB, EviCastKV obtains 21.88\% temporal Accuracy and 14.52
semantic ROUGE-L, versus 3.12\% and 12.31 for LOOK-M.
Paired gains are $18.75$ percentage points, interval $[10.42,27.08]$,
and $2.21$ ROUGE-L points, interval $[0.80,3.62]$. Mean request
time falls from 24.36 to 14.95~s (38.6\%). These findings concern
the local byte-capped port; the aggregate results do not identify
merging, parsing, or another implementation detail as the cause of
its degradation.

At 40~MiB (Table~\ref{tab:mile40}), the ROUGE-L gain over LOOK-M is
$1.36$, interval $[0.15,2.55]$, but Accuracy differs by $-2.08$
points, interval $[-7.29,3.12]$. Against Top-K, paired correctness
matches and the ROUGE-L difference is unresolved ($-0.04$, interval
$[-0.25,0.17]$). Across five capacities, mean request time is
38.6--58.4\% lower than LOOK-M and 5.6--7.5\% lower than Top-K.
The primary FullKV temporal Accuracy of 22.92\% and limited subtask
coverage constrain broader reasoning claims. Smaller-budget tests
do not establish joint quality noninferiority.

\begin{table}[t]
\centering
\caption{MileBench exact selection: LLaVA-1.5-7B, FP16, 64-token
protocol. Memory is mean active K/V backing, not audited decode
peak. Each quality metric has 96 examples; E2E is the mean over
192 requests. R-L denotes ROUGE-L $\times100$. Bold marks column-best
point estimates.}
\label{tab:mile40}
\begingroup
\footnotesize
\setlength{\tabcolsep}{2pt}
\begin{tabular*}{\columnwidth}{@{\extracolsep{\fill}}lrrrrr@{}}
\toprule
& \multicolumn{2}{c}{K/V backing (MiB)}
& \multicolumn{2}{c}{Quality} &\\
\cmidrule(lr){2-3}\cmidrule(lr){4-5}
Method & Initial & Final & Acc. (\%) & R-L & E2E (s)\\
\midrule
Prompt Top-K
  & 40.00
  & 71.50
  & 22.92
  & \textbf{15.88}
  & 18.03\\
LOOK-M 
  & 40.00
  & 71.50
  & \textbf{25.00}
  & 14.48
  & 34.22\\
EviCastKV
  & 40.00
  & 71.50
  & 22.92
  & 15.85
  & \textbf{16.71}\\
\bottomrule
\end{tabular*}
\endgroup
\end{table}

From a common 40-MiB initial cap, continuation adds 7.875~MiB on COCO and 31.50~MiB on MileBench.
Budgets match methods within each checkpoint, not final storage across checkpoints; shared-source accounting remains distinct.

\subsection{Runtime of the Exact-Selection Path}
\label{sec:eval_runtime}

Only 0.1724~s of the 4.0776~s COCO request-time reduction over VL-Cache at 40~MiB lies in decoding. On MileBench at 40~MiB, planning/construction costs 10.50~s for EviCastKV, 11.63~s for
Top-K, and 26.36~s for LOOK-M, while decode means are 2.77--2.80~s. Complete-request gains therefore coexist with substantial compression overhead. These measurements do not isolate forecasting, graph, or routing contributions, and equal K/V bytes rule out smaller cache volume as their explanation. 


\begin{table*}[t]
\centering
\caption{EviCastKV shared-source operating points on 256 COCO images with LLaVA-1.5-7B, BF16, and
128 predictions. Decode-owned storage includes accounted metadata; the audited peak also includes observed reconstruction/append overlap. Initial/final columns are means; peak mean/max aggregate per-request audited peaks. This is not the primary study's K/V-only boundary maximum or whole-request memory. Realized byte budgets differ. The last column gives familywise 95\% intervals for EviCastKV minus the indicated reference in CIDEr.}
\label{tab:storage-operating}
\begingroup
\small
\setlength{\tabcolsep}{3pt}
\begin{tabular*}{\textwidth}{@{\extracolsep{\fill}}lrrcrrrl@{}}
\toprule
& \multicolumn{3}{c}{Decode-owned storage (MiB)}
& & \multicolumn{2}{c}{Request time (s)} &\\
\cmidrule(lr){2-4}\cmidrule(lr){6-7}
Method
& Initial
& Final
& \shortstack{Audited peak\\mean/max}
& CIDEr
& p50
& p95
& \shortstack{Familywise 95\% CI\\for $\Delta$CIDEr}\\
\midrule
EviCastKV
  & \textbf{41.14}
  & \textbf{73.22}
  & \textbf{83.52/84.24}
  & 119.77
  & 6.55
  & 6.72
  & ---\\
Prompt Top-K (A)
  & 50.73
  & 82.80
  & 92.83/92.95
  & 118.33
  & 7.50
  & 7.66
  & $[-2.713,5.712]$\\
PrefixKV (A)
  & 45.70
  & 77.78
  & 86.01/86.56
  & 103.10
  & 7.38
  & 7.54
  & $[4.677,27.616]$\\
VL-Cache (A)
  & 53.31
  & 85.39
  & 94.39/94.64
  & \textbf{122.16}
  & 13.25
  & 17.66
  & $[-7.805,2.533]$\\
LOOK-M (A)
  & 50.72
  & 82.80
  & 92.82/92.83
  & 120.96
  & 48.86
  & 51.59
  & $[-6.670,4.182]$\\
\bottomrule
\end{tabular*}
\endgroup
\end{table*}

\subsection{Shared-Source Storage and Reconstruction}
\label{sec:eval_storage}

\textbf{Protocol:}
EviCastKV's shared-source configuration uses BF16 LLaVA-1.5-7B on 256 inspected COCO images, with
603 prompt positions and 128 fixed greedy predictions. Scoring ends at the first terminal token;
systems execution performs 127 post-prefill forwards.
Each request uses one A100. (A) denotes precision/storage adaptation
\cite{wang2025prefixkv,tu2025vlcache,wan2024lookm}.

\textbf{Storage and quality:}
EviCastKV has the smallest initial, final, and mean audited
decode peak in Table~\ref{tab:storage-operating}.
Relative to Top-K, these decrease by 18.89\%, 11.57\%,
and 10.03\%. Its 119.77 CIDEr exceeds Top-K's point
estimate but is below VL-Cache and LOOK-M. Ten thousand
paired image bootstraps recompute corpus CIDEr, with
Bonferroni intervals for five contrasts. Only the
PrefixKV-adaptation gain of 16.67 has a positive familywise
interval, $[4.677,27.616]$; its 42/256 output-limit-hit
captions remain scored. Quality preservation against
Top-K, VL-Cache, and LOOK-M is not established. Table~\ref{tab:storage-operating} reports all five familywise quality intervals, p95 request times, and the mean and maximum of the per-request audited decode peaks.

\textbf{Timing and validation:}
All 1,536 audited/unaudited method--image pairs match token IDs and storage endpoints; six canaries match all 128 exported logits bitwise. 
Timing follows five warmups per method per worker and randomized method order.
Each uninstrumented request recomputes preprocessing, prefill, construction, replay setup, and decoding, including logits export, detokenization, and retirement. 

EviCastKV's median/p95 is 6.55/6.72~s, below the compressed adaptations 
. A separate six-image, 20-repeat experiment reports mean-time reductions of 12.35\%, 11.07\%, 22.13\%, and 86.14\% against Top-K, PrefixKV, VL-Cache, and LOOK-M adaptations, using four-contrast-adjusted nested paired intervals. These ratios are not full-cohort median reductions. Replay engages for 256/256 EviCastKV requests and 0/256 requests for each compressed adaptation, so latency combines representation, layout, and execution effects.

\section{Related Work}
\label{sec:related_work}

\textbf{KV selection and anticipation: } H$_2$O retains heavy-hitter and recent tokens; StreamingLLM protects attention sinks; SnapKV uses prompt-window observations; and
PyramidKV varies layer-wise capacity \cite{zhang2023h2o,xiao2024streaming,li2024snapkv,cai2024pyramidkv}. Expected Attention estimates future attention from query distributions~\cite{devoto2025expectedattention}, while Attention
Matching constructs compact keys and values to approximate attention outputs and mass~\cite{attentionmatching2026}. EviCastKV combines prompt-conditioned selection with value-sensitive neighborhood and modality coverage, then packs selected original K/V entries under a physical-byte constraint. 

\textbf{Multimodal cache compression: } LOOK-M combines text-prior selection with compensatory merging; PrefixKV allocates layer-wise importance-ranked prefixes; and VL-Cache combines modality-aware scoring with sparsity-adaptive budgets~\cite{wan2024lookm,wang2025prefixkv,tu2025vlcache}.
AirCache estimates inter-modal relevance using an elite observation window and adaptive budgets~\cite{huang2025aircache}. LightKV uses prompt-guided cross-modal message passing to aggregate visual representations during prefill~\cite{chen2026lightkv}. EviCastKV's exact-retention configuration instead uses sparse relationships to guide retention of original states,
without merging or reconstructing retained values.

\textbf{Quantization and latent representations:} KIVI quantizes keys per channel and values per token \cite{liu2024kivi}. Palu decomposes projections, caches low-rank intermediate states, and reconstructs K/V during execution \cite{chang2025palu}. EviCastKV's shared-source configuration stores the frozen model's normalized attention input with source quantization, exact protected overlays, and lossless coding of its own generated history. Unlike low-rank factorization, it retains the original projection matrices. The contribution concerns this representation and its execution contracts, rather than reconstruction from intermediate state in isolation.

\textbf{Distributed workflows and dataframe execution.}
Radical-Cylon integrates Cylon dataframe tasks with RADICAL-Pilot
for execution across heterogeneous resources~\cite{sarker2024radical}.
Deep RC combines data engineering and distributed deep learning
in workflows spanning preprocessing, training, and
postprocessing~\cite{osti_10634837}.
CylonFlow integrates Cylon's dataframe execution into Dask and Ray
through stateful execution and modular communication
backends~\cite{perera2023supercharging, perera2023depth, staylor2025combining, shan2022hybrid}.
These systems improve efficiency through workflow orchestration
and distributed data processing. EviCastKV addresses a complementary
level of efficiency through the selection and physical representation
of KV states within multimodal inference requests.

\textbf{Efficient execution:} DuoAttention assigns full and bounded caches to retrieval and streaming heads, respectively~\cite{xiao2025duoattention}. EviCastKV instead routes \emph{planning effort} between graph selection and prompt ranking. FlashAttention and FlashAttention-2 optimize attention IO and work partitioning \cite{dao2022flash,dao2023flash2}; PagedAttention improves physical allocation and sharing~\cite{kwon2023paged}. AAFLOW+ concerns distributed KV-state operator abstraction, whereas OpRAG concerns resource-aware orchestration that helps hybrid retrieval effectively~\cite{sarker2026oprag, sarker2026aaflow+}. These mechanisms address different execution boundaries from evidence selection and source encoding. Our eager-runtime measurements do not establish fused-kernel integration or serving-throughput gains.
\section{Discussion and Future Work}
\label{sec:discussion}
\label{sec:limitations}

\textbf{Evidence-aware selection:}
At matched capacity, EviCastKV improves COCO CIDEr by
3.64 over local VL-Cache at 40~MiB and tight-budget
MileBench quality over local LOOK-M. These findings
support the combined approach: prompt-conditioned,
value-sensitive coverage guides retention, while routed
planning and position-preserving construction translate
selection into physically feasible execution.

\textbf{Physical storage:}
Capacity reconciliation accounts for mandatory entries and padding rather than relying on nominal retention. EviCastKV also addresses representation cost through its shared-source configuration, combining exact protection with encoded source storage and layer-local reconstruction. Relative to the measured Top-K adaptation, initial storage and mean audited decode peak decrease by 18.89\% and 10.03\%, respectively. These operating-point savings demonstrate a representation opportunity, with accompanying quality trade-offs reported
in Table~\ref{tab:storage-operating}.

\textbf{Complete-request efficiency:}
EviCastKV reduces mean COCO request time by 19.65\%
against local VL-Cache and MileBench time by 5.6--7.5\%
against Top-K across tested budgets. Similar decode times
at matched capacity highlight planning and construction
as important optimization targets. Effective compression
must therefore consider evidence selection, materialization,
and complete execution cost together.

\textbf{Future directions:}
Priorities are routing and utility ablations on independent workloads and graph-guided shared-source protection at common realized-byte budgets, separating representation from replay effects. Deployment work will audit prefill/construction memory, maintain budgets during long generation, and explore GQA-aware encoding, fused or paged attention, and batching.
\section{Conclusion}
\label{sec:conclusion}

We presented EviCastKV, a training-free approach to query-conditioned multimodal cache construction. Its exact-retention configuration combines value-sensitive
inter-modal coverage, routed planning, and position-preserving
packing and improves COCO quality over local VL-Cache with 19.65\% lower mean request time,
and tight-budget MileBench quality over local LOOK-M
with 38.6\% lower mean request time.

The shared-source configuration complements these findings
through quantized source storage, protected exact entries,
and layer-local reconstruction. Its measured operating point
reduces decode-owned storage relative to the tested compressed
adaptations, with the corresponding quality and execution
trade-offs reported explicitly. Together, the configurations
show the importance of considering evidence prioritization,
physical representation, and complete request cost jointly.
Future work will isolate component contributions and evaluate
graph-guided source protection on independent workloads.

{
    \small
    \bibliographystyle{ieeenat_fullname}
    \bibliography{main}
}


\end{document}